\documentclass{article}
\usepackage{array}

\begin{document}
\setcounter{page}{0}

\thispagestyle{empty}
\begin{center}
LaRIA~: Laboratoire de Recherche en Informatique d'Amiens\\
Universit\'e de Picardie Jules Verne -- CNRS FRE 2733\\
33, rue Saint Leu, 80039 Amiens cedex 01, France\\
Tel : (+33)[0]3 22 82 88 77\\
Fax : (+33)[0]03 22 82 54 12\\
\underline{http://www.laria.u-picardie.fr}
\end{center}

\vspace{7cm}

\begin{center}
\parbox[t][5.9cm][t]{10cm}
{\center

{\bf Sudo-Lyndon

\medskip

G. Richomme$^{\rm a}$
}

\bigskip

\textbf{{L}aRIA \textbf{R}ESEARCH \textbf{R}EPORT~: LRR 2007-03}\\
(February 2007)
}
\end{center}

\vfill

\hrule depth 1pt \relax

\medskip

\noindent
$^a$ LaRIA, Universit\'e de Picardie Jules Verne, gwenael.richomme@u-picardie.fr\\

\vspace{-2cm}
\pagebreak

\title{Sudo-Lyndon}
\author{Gw\'en\"el Richomme\\
LaRIA, Universit\'e de Picardie Jules Verne\\
33 Rue Saint Leu,
80039 Amiens cedex 01, France\\
{gwenael.richomme@u-picardie.fr}\\
{http://www.laria.u-picardie.fr/~richomme}
}
\date{\today}
\maketitle

\begin{abstract}
Based on Lyndon words, a new Sudoku-like puzzle is presented and some
relative theoretical questions are proposed.
\end{abstract}

\section{Introduction}

Lyndon words are basic tools in Combinatorics on Words (see for
instance \cite{Lothaire1983,Lothaire2002}) and appear in many problems
that can be expressed using words (see for instance
\cite{Chemillier2004,DieudonnePetit2006,Smyth2003}).  While preparing
a course, I thought about the puzzle presented in
Section~\ref{thePuzzle} for a pleasant first contact with Lyndon words
(see also Section~\ref{variants} for some possible variants).  I
propose to nickname \textit{Sudo-Lyndon} this
puzzle since, as in the now famous (who did not hear about it?) Sudoku
game, also known as Number Place puzzle, a grid has to be filled from
partial informations, and only a unique solution is expected as a result
(may be a better name should be \textit{Lyndon place}).

\section{\label{thePuzzle}The puzzle}

For our purpose a \textit{word} is nothing else than a finite
\textit{non-empty} sequence of letters: here the meaning of the word
does not care.

A \textit{Lyndon word} is a word smaller, with respect to the
lexicographic order, than all its suffixes except itself (See for
instance \cite{Lothaire1983} for equivalent definitions and
properties). For example, when considering the usual ordering of the
letters, one can verify that \textit{cocoon} or \textit{acacias} are
Lyndon words while \textit{bananas}, \textit{acacia},
\textit{anagram} or \textit{eighteen} are not.

Here we will play only with two letters $a$ and $b$, but we
will simultaneously consider Lyndon words over $\{a < b\}$ (that is over
the alphabet $\{a, b\}$ with $a < b$) and over $\{b < a \}$. The
shortest Lyndon words over $\{ a < b\}$ are $a$, $b$, $ab$, $aab$,
$abb$, $aaab$, $aabb$, $abbb$, $aaaab$, $aaabb$, $aabab$, $aabbb$,
$ababb$ and $abbbb$. The shortest ones over $\{b < a\}$ are of course
obtained replacing each $a$ by a $b$ and each $b$ by an $a$. Lyndon
words of length 6 over $\{b < a\}$ are $bbbbba$, $bbbbaa$, $bbbaaa$,
$bbbaba$, $bbabaa$, $bbaaba$, $bbaaaa$, $babaaa$, $baaaaa$.

\medskip

The \textit{aim of the puzzle} is to fill each cell of a grid with a
letter $a$ or $b$ so that each row read from left to right and each
column read top-down yields a Lyndon word over $\{a < b\}$ or
over $\{b < a\}$. For each initial grid, the set of predetermined
cells is defined in such a way that one and only one correct solution
exists.
Let us give an example with its solution~:

\begin{center}
\begin{minipage}{4cm}
\center
Puzzle\\[+.2cm]
\setlength\extrarowheight{3pt}
\begin{tabular}{|l|l|l|l|}
\hline
 &  &  & \\
\hline
~~ & a & b & ~~ \\
\hline
& b & a &  \\
\hline
 & & & \\
\hline
\end{tabular}
\end{minipage}
\begin{minipage}{4cm}
\center
Solution\\[+.2cm]
\setlength\extrarowheight{3pt}
\begin{tabular}{|l|l|l|l|}
\hline
a & a & b & b \\
\hline
a & a & b & b \\
\hline
b & b & a & a \\
\hline
b & b & a & a \\
\hline
\end{tabular}
\end{minipage}

\end{center}

Here follow two other examples (solutions can be found on my home page)~:
\begin{center}
\begin{minipage}{4cm}
\center
Puzzle 1\\[+.2cm]
\setlength\extrarowheight{3pt}
\begin{tabular}{|l|l|l|l|l|}
\hline
a & & & & \\
\hline
& b & b & & \\
\hline
& a & b & & a \\
\hline
a & & a & & \\
\hline
& & & b & \\
\hline
\end{tabular}
\end{minipage}
\begin{minipage}{6cm}
\center
Puzzle 2\\[+.2cm]
\setlength\extrarowheight{3pt}
\begin{tabular}{|l|l|l|l|l|l|l|l|l|}
\hline
   &   &   &   &   &   &   & a &    \\ 
\hline
   & a & b &   &   & a & a & b &    \\ 
\hline
   & b &   & a &   &   &   &   &    \\ 
\hline
 b & a &   & b &   &   &   &   &    \\ 
\hline
 a &   &   &   & a &   &   &   &    \\ 
\hline
 a & a &   & b & a & a & b &   &    \\ 
\hline
   &   &   & a & a &   & b &   & a  \\ 
\hline
   &   &   &   &   &   &   &   &    \\ 
\hline
\end{tabular}
\end{minipage}
\end{center}

\section{Educational matters}

As explained in the introduction, my aim when designing my first grid
was educational. Let me narrate my experience with it. I tested it
with some students (Puzzle 1 was made as an introductory exercise and
Puzzle 2 was given as homework). Most of the students managed to find
the solution, often empirically, sometimes bactracking.  We then
discussed on the way to obtain the solution the most directly as
possible. Quickly it was observed by students that the first and last
letters of a Lyndon word must be different. I then introduced the fact
that ``{\it a Lyndon word is unbordered\,}'' (a word $w$ is unbordered
if the only word which is both prefix and suffix of $w$ is $w$ itself
$w$) and its corollary ``{\it a Lyndon word is primitive\,}'' (that
is, it is not a power of a strictly smaller word). We deduced basic
rules to fill a (enough large) grid, as for instance:\\ \centerline{\textit{Rule
1}: $?a\ldots b? \to aa\ldots bb$}\\ \centerline{\textit{Rule 2}:
$ab\ldots ?? \to ab\ldots bb$}

Symbol $?$ in rules denotes an unknown letter. Rule 1 means that if we
do not know the first and penultimate letter of a Lyndon word, if the
second letter is {\it a} and if the last letter is a {\it b}, then the
Lyndon word must start with $aa$ and must end with $bb$. Rule 2 means that 
any Lyndon word starting with $ab$ must end with $bb$.

One can observe that Puzzle 1 can be solved using only the
unborderedness of Lyndon words. Hence Puzzle 2 should be a better
introduction. It allows to remark that if a Lyndon word $w$ starts with
$a^nb$ for a non-zero integer $n$, then $a^{n+1}$ is not a factor of
$w$ (a factor of a word is a subsequence of consecutive letters).

\medskip

To end with educational matters, I would like to notice that before
starting the exercise with students, I just defined Lyndon words and
give few examples. While discussing the solution, I mentioned that
another approach could have been to enumerate Lyndon words of length 5
(only 6 such words exist over $\{a < b\}$) and to try to put them in the
grid. To explain it, I gave the lists of Lyndon words over $\{a < b\}$
for each length from 1 to 5. This lead a student to ask for the number
of Lyndon words for each length (see \cite{Lothaire1983} for an answer).

\section{Some theoretical questions}

All grids presented here were handmade. 
Hence a natural question is:

\noindent
\textit{Problem 1}: How to generate (effectively) a Sudo-Lyndon puzzle?

To answer this question, one would certainly need to know the
structure of possible solutions of a Sudo-Lyndon puzzle. Without any
information of this kind, a basic idea is to generate a candidate grid
until it has a unique solution. But for this we need an answer to the
following second natural question:

\noindent
\textit{Problem 2}: Given a grid partially filled with letters $a$ and
$b$, how to know (effectively) if there exists a unique solution?

Of course, for each of the previous questions, we vould like to know
its complexity class. In particular is Problem 2 NP-complete as is the
similar question for Sudoku game \cite{YatoSeta2002}?

\medskip

We observe that it can be determined in linear time with respect to
the number of cells whether a grid filled with letters $a$ and $b$ is
a possible solution of a puzzle, that is, whether each row and column
yields a Lyndon word. This is an immediate consequence of the
existence of a linear time algorithm to check whether a word is a
Lyndon word (see \cite[chapter 1]{Lothaire2005} for references).

\medskip

 A sub-question to Problem~2 concerns words or more precisely partial
 words as defined by J.~Berstel and L.~Boasson \cite{BerstelBoasson1999} and
 studied in depth by F.~Blanchet-Sadri (see for instance
\cite{Blanchet-Sadri2005}). A \textit{partial word} is a
 word with \textit{holes}, that is, with positions where letters are
 undetermined. Each row/column in a Sudo-Lyndon puzzle is
 a partial word. Hence:

\noindent
\textit{Problem 3}: given a partial word, can we replace a letter in
such a way that the result yields a Lyndon word?

This third problem is being studied by Blanchet-Sadri and Davis
\cite{Blanchet-SadriDavis}.

\medskip

\begin{minipage}{8cm}
The grid on the right shows that a positive answer to Problem~3 is not
sufficient to solve Problem~2 since in the unique hole an occurrence
of the letter $b$ is needed to have a horizontal Lyndon word whereas
an occurrence of the letter $a$ is needed to have a vertical Lyndon
word.
\end{minipage}
\begin{minipage}{4cm}
\center
No solution!\\[+.2cm]
\setlength\extrarowheight{3pt}
\begin{tabular}{|l|l|l|l|l|}
\hline
a & a & b & b & b\\
\hline
a & a & b & a & b\\
\hline
a & a & b & b & b \\
\hline
a & b & a & & b\\
\hline
b & b & a & a & a\\
\hline
\end{tabular}
\end{minipage}

\medskip

To end this section, we consider the question, connected to Problems 1
and 2, of the enumeration of all the possible solutions of the
puzzle. The scheme below shows that the number of such grids grows
exponentially with the number of rows and columns. Indeed in the
scheme each $*$ symbol can be replaced independently by an $a$ or a $b$
in order to obtain a possible solution of a puzzle. Hence
$2^{(\lfloor \frac{n}{2} \rfloor - 1) (\lfloor \frac{m}{2}\rfloor -1)}$
different solutions of the puzzle can be available with this scheme.

\medskip

\begin{tabular}{c|c|c|c|c|c|c|c|c|}
\multicolumn{1}{c}{}   
& \multicolumn{1}{c}{1} 
   & \multicolumn{1}{c}{2}  
   & \multicolumn{1}{c}{\ldots}
   & \multicolumn{1}{c}{\small $\lfloor \frac{m}{2} \rfloor$}
   & \multicolumn{1}{c}{\small $\lfloor \frac{m}{2} \rfloor + 1$}
%   & \multicolumn{1}{c}{\small $\lfloor \frac{m}{2} \rfloor + 2$}   
   & \multicolumn{1}{c}{\ldots}
   & \multicolumn{1}{c}{m-1}
   & \multicolumn{1}{c}{m}
\\ 
\cline{2-9}
1 & a & a & \ldots & a & b & \ldots & b & b \\
\cline{2-9}
2 & a & a & \ldots & a & b & \ldots & b & b \\
\cline{2-9}
& \vdots & \vdots & $\ddots$ & \vdots & \vdots & 
$\ddots$ & \vdots & \vdots \\
\cline{2-9}
{\small $\lfloor \frac{n}{2} \rfloor$} & a & a & \ldots & a & b & $\ddots$ & b & b \\
\cline{2-9}
{\small $\lfloor \frac{n}{2} \rfloor + 1$} & b & b & \ldots & b & $*$ & $\ddots$ & $*$ & a \\
\cline{2-9}
& \vdots & \vdots & $\ddots$ & \vdots & \vdots &  
$\ddots$ & \vdots & \vdots \\
\cline{2-9}
n-1 & b & b & \ldots & b & $*$ &  \ldots & $*$ & a \\
\cline{2-9}
n & b & b & \ldots & b & a & \ldots & a & a \\
\cline{2-9}
\end{tabular}

\section{\label{variants}Variants}

When constructing the grids presented here, I was annoying by the
feeling that two much cells were filled in the initial grid of the
puzzle. This leads to the problems:

\noindent
\textit{Problem~4}: given a puzzle for which we know there exists a unique
solution, can we determine if it is minimal in the sense that no
letter in the grid can be deleted without losing the uniqueness of the
solution?

\noindent
\textit{Problem~5}: given integers $n, m$, what is the minimal number
$f(n, m)$ such that there exists at least one grid that starts with
$f(n, m)$ initial values and has a unique solution?

In the particular case of a word, one can see that the ratio of this
minimal amount over the length of a word tends to 0 when the length of
the word tends to infinity. Indeed the partial word (each hole is
indicated by a ? character)

\vspace{-0.3cm}

$$ab^p?a?^{2p+2}[a?^{2p+3}]^p$$ 

\vspace{-0.2cm}

is of length $2p^2+7p+5$ and contains
only $2p+2$ known letters, and the Lyndon word
$ab^{p+1}ab^{2p+2}[ab^{2p+3}]^p$ is the unique solution to this one
dimension puzzle.

\medskip

Finally, to deal with the problem of having too much information, I
have thought about the following variants.  For each variant, the aim
is still to fill each cell of a grid with a letter $a$ or $b$ so that
each row read from left to right and each column read top-down
yields a Lyndon word over $\{a < b\}$ or over $\{b < a\}$.

\begin{description}
\labelsep0cm
\partopsep0cm
\parskip0cm
\topsep0cm
\itemsep0cm
\item{\textit{Variant 1:}} For each row/column, the number of occurrences of
the letter $a$ is given. 

\item{\textit{Variant 2:}} As for variant~1, but the value of some cells
are also given.

\item{\textit{Variant 3:}}(As for the original Sudoku puzzle), the grid is
divided into subgrids. For each subgrid, we consider the word obtained
reading the rows of the subgrid from the top one and from left to right
(One can prefer to concatenate columns and so naturally other variant
is to consider simultaneously the two possibilities). In this variant
the global grid should be filled in such a way a Lyndon word is
written on each row, column and subgrid.

\item{\textit{Variant 4:}} As for variant 3, but moreover some cells can be
initially filled with the * character meaning that in the final
solution the value of the cell can be equally the letter a and b. Of
course, as for the initial puzzle, there is only one manner to fill a
cell (except those marked with a *) for which the value is not known at
the beginning.
\end{description}

%Here follow some examples:

\begin{center}
\begin{minipage}{6cm}
\center
Variant 1\\[+.2cm]
\setlength\extrarowheight{3pt}
\begin{tabular}{c|c|c|c|c|c|c|}
\multicolumn{1}{c}{}   & 
     \multicolumn{1}{c}{4} 
   & \multicolumn{1}{c}{5}  
   & \multicolumn{1}{c}{3}
   & \multicolumn{1}{c}{3}
   & \multicolumn{1}{c}{2}
   & \multicolumn{1}{c}{2}   \\ 
\cline{2-7}
2  &  &   &   &   &  &    \\ 
\cline{2-7}
2  &  &   &   &   &  &    \\ 
\cline{2-7}
4  &  &   &   &   &  &    \\ 
\cline{2-7}
2  &  &   &   &   &  &    \\ 
\cline{2-7}
5  &  &   &   &   &  &    \\ 
\cline{2-7}
4  &  &   &   &   &  &    \\ 
\cline{2-7}
\end{tabular}
\end{minipage}
\begin{minipage}{6cm}
\center
Variant 2\\[+.2cm]
\setlength\extrarowheight{3pt}
\begin{tabular}{c|c|c|c|c|c|}
\multicolumn{1}{c}{}   & 
     \multicolumn{1}{c}{3} 
   & \multicolumn{1}{c}{4}  
   & \multicolumn{1}{c}{2}
   & \multicolumn{1}{c}{2}
   & \multicolumn{1}{c}{2}
   \\ 
\cline{2-6}
2  & a &   &   &   &     \\ 
\cline{2-6}
3  &  &   &   &   &      \\ 
\cline{2-6}
2  &  &   &   &   &      \\ 
\cline{2-6}
3  &  &   &   &   &      \\ 
\cline{2-6}
3  &  &   &   &   &      \\ 
\cline{2-6}
\end{tabular}
\end{minipage}
\end{center}

\noindent
\begin{minipage}{5 cm}
\center
Variant 3\\[+.2cm]
\setlength\extrarowheight{3pt}
\begin{tabular}{|l|l|l||l|l|l|}
\hline
~~ & ~~ &  ~~ & ~~ & ~~ & ~~ \\
\hline
 & a &  & & b & \\
\hline
 & & & & &b\\
\hline
 & & & & &\\[-.5cm]
\hline
 & & & & & \\
\hline
 & b & a & a & a & \\
\hline
 & & a & & & \\
\hline
\end{tabular}
\end{minipage}
\begin{minipage}{6cm}
\center
Variant 4 \\[+.2cm]
\setlength\extrarowheight{3pt}
\begin{tabular}{|l|l|l||l|l|l||l|l|l|}
\hline
~~ & ~~ &  b & a & ~~ & ~~ & a & ~~ & ~~ \\
\hline
 & a &   & b & b & & b &  b & \\
\hline
 &  &   &  & & & &  & \\
\hline
 &&&&& & & &\\[-.5cm]
\hline
 &  a & b & &  & b  & &   & a\\
\hline
 & b &  & & b  & $*$ & &   & \\
\hline
 &  &  &  & $*$ & &  & $*$  & \\
\hline
 &&&&& &  & &\\[-.5cm]
\hline
  & &  & & b & $*$ & &   & \\
\hline
  & b & $*$ &  & a & $*$ &  & a  & \\
\hline
  & &  & &  & & &   & \\
\hline
\end{tabular}
\end{minipage}

\section{Conclusion}

We have already mentioned that Puzzle 1 could be solved using only
unborderdness. This kind of puzzle can of course be generalized to
larger grid, but it can also open the question to find interesting
puzzles based on other properties of words. For instance all Lyndon
words in the grid below do not contain the words $aaa$ and $bbb$ as
factors.

\begin{center}
\begin{minipage}{6cm}
\setlength\extrarowheight{3pt}
\begin{tabular}{|l|l|l|l|l|l|l|l|l|}
\hline
a & b & ~~ & ~~ & ~~ & ~~ & a & ~~ & ~~ \\
\hline
 & a &   &  & b & &  &   & \\
\hline
 &  &   &  & & b & &  & \\
\hline
 &  & & &  &  & &   & \\
\hline
 &  &  & &   & & & a  & \\
\hline
 &  &  &  & & &  &  & \\
\hline
  & &  & &  & & &   & \\
\hline
  &  &  &  &  & &  & b  & \\
\hline
  & &  & &  & & &   & \\
\hline
\end{tabular}
\end{minipage}
\end{center}

%\medskip

~ \hfill I hope that you get fun playing Sudo-Lyndon.

\bigskip

\noindent
\textbf{Acknowledgements}. Many thanks to my beta-testers students, to
Richard Groult and especially to Patrice Séébold who encourages me to
widen the communication of the present puzzle.

\end{document}